\documentclass[fleqn,twoside]{article}%
\usepackage{amsmath}
\usepackage{amsfonts}
\usepackage{amssymb}
\usepackage{graphicx}
\usepackage{cite}
\def\be{\begin{equation}}
\def\ee{\end{equation}}
\def\bi{\bibitem}
\begin{document}
\title{Bianchi Type-II Cosmological Model with Viscous Fluid.}
\author{A. Banerjee$^1$, S. B. Duttachoudhury and Abhik Kumar Sanyal$^2$}
\maketitle
\noindent
\begin{center}
\noindent
Department of Physics, Jadavpur University, Calcutta 700 032, India.\\
\end{center}
\footnotetext{\noindent
Electronic address:\\
\noindent
$^1$ asit@juphys.ernet.in\\
$^2$ sanyal\_ ak@yahoo.com;
Present address: Dept. of Physics, Jangipur College, India - 742213.}
\noindent
\abstract{A spatially homogeneous and locally rotationally symmetric Bianchi type-II cosmological model under the influence of both shear and bulk viscosity has been studied. Exact solutions are obtained with a barotropic equation of state ($p = \epsilon\rho$) and considering the linear relationships between $\rho$, $\theta^2$, and $\sigma^2$, which represent the fluid density, the expansion, and the shear scalars, respectively. Special cases with vanishing bulk viscosity coefficients and with the perfect fluid in the absence of viscosity have also been studied. The formal appearance of the solutions is the same for both the viscous as well as the perfect fluids. The difference is only in choosing a constant parameter which appears in the solutions. In the cases of either a fluid with bulk viscosity alone or a perfect fluid, the barotropic equation of state is no longer an additional assumption to be imposed; rather it follows directly from the field equations.}

\maketitle
\flushbottom
\section{Introduction:}

In the recent years the introduction of viscosity in the fluid content has been found to explain successfully many a physical features in the dynamics of homogeneous cosmological models. Since viscosity counteracts the cosmological collapse a different picture at the initial state of cosmological evolution may appear due to the dissipative process caused by viscosity. \\

Attempts have been made by several authors to find exact solutions of Einstein's field equations by considering viscous fluid in isotropic as well as anisotropic cosmological models. Murphy \cite{1} has given the solution for the flat cosmological model of the Friedmann type taking into account the effect of bulk viscosity only, which has lately been extended by Banerjee and Santos \cite{2}. Belinskii and Khalatnikov, considering a qualitative analysis of isotropic \cite{3} and anisotropic \cite{4} Bianchi type-I cosmological models with viscous fluid discussed only the asymptotic behaviors in different cases. In a previous paper \cite{5} the present authors gave some exact solutions for viscous fluid cosmological models of Bianchi type-I and discussed the role of viscosity in determining the nature of the singularity. Some such solutions already exist in the literature (Banerjee and Santos \cite{6}) with a larger number of restrictions on the nature of the fluid.\\

In the present paper we have considered a locally rotationally symmetric (LRS) model of spatially homogeneous Bianchi type-II cosmology, where a barotropic equation of state for the fluid holds. Two additional assumptions are made in order to make the number of unknowns compatible with the number of available independent equations. One of them, viz., ${\sigma^2\over \theta^2} =$ const, was previously utilized by some authors like Collins \cite{7} to obtain exact solutions for perfect fluid. Here, we have found that the matter density is infinitely large near the initial singularity, where the proper volume goes to zero. Further we have shown that the effect of viscosity is more important at the initial epoch and the expansion and shear scalars both approach negligible values asymptotically at the same rate. It has also been shown that the effect of viscosity gives rise to a large accumulation of entropy during the final stage of evolution, which may be
significant in connection with the existence of the large entropy per baryon in the present state of the universe.\\

In Section 2, we have solved Einstein's field equation considering a barotropic equation of state ($p=\epsilon\rho$) along with the two additional assumptions, viz., ${\sigma^2\over \theta^2} =$ const and ${\rho\over \theta^2} =$ const, where $\sigma$, $\theta$ and $\rho$ stand for shear, expansion, and mass density, respectively. In Section 3 we have considered two special cases, firstly for a fluid having only bulk viscosity and secondly for a perfect fluid. In both cases the barotropic equation of state follows from the field equations directly and as such is not introduced as an assumption. The stiff equation of state $p = \rho$ is found to be inadmissible in the perfect fluid case. The nature of the solutions in all the cases is found to be formally the same with, however, a difference in the magnitude of a constant. It is found that the Hawking-Penrose energy conditions demand certain restrictions on the magnitude of the parameter 2 appearing in the solutions.

\section{General Solutions of Einstein's field equations:}

The LRS metric for the spatially homogeneous Bianchi type-II model \cite{8,9} is

\be\label{2.1} ds^2 = -dt^2 + S^2(dx + z dy)^2 + R^2 (dy^2 + dz^2),\ee
where $R = R(t)$ and $S = S(t)$. The energy-momentum tensor of the viscous fluid \cite{10} is given by,

\be\label{2.2} T_{ij} = (\rho + \bar p)v_i v_j + \bar p g_{ij} - \eta \mu_{ij},\ee
with
\be\label{2.3} \bar p = p - \left(\zeta -{2\over 3}\eta\right){v^a}_{;a},\ee
and
\be\label{2.4} \mu_{ij} = v_{i;j} + v_{j;i} + v_i v^a v_{j;a} + v_j v^a v_{i;a}.\ee
In the above equations $\zeta$ and $\eta$ stand for the bulk and shear viscosity coefficients, respectively, $\rho$ and $p$ are the mass density and isotropic pressure,
respectively, $\bar p$ is the effective pressure, and $v_i$ represents the four-velocity,
so that
\be\label{2.5} v_i v^j = -1.\ee
In the system of units $8\pi G = C = 1$, Einstein's field equations can be written as:
\be\label{2.6} {R_i}^j - {1\over 2} {\delta_i}^j R = - {T_i}^j.\ee
Using a comoving coordinate system, so that $v^i = {\delta_0}^i$ the explicit forms of the field equations \eqref{2.6} can be written, in view of equations \eqref{2.1} - \eqref{2.5}, as

\be\label{2.7} 2{\dot R\over R}{\dot S\over S} + {\dot R^2\over R^2} - {1\over 4}{S^2\over R^4} = \rho.\ee
\be\label{2.8} {\ddot S\over S} + {\ddot R\over R} + {\dot R\over R}{\dot S\over S} + {1\over 4}{S^2\over R^4} = -\bar p + 2\eta {\dot R\over R}.   \ee
\be\label{2.9} 2{\ddot R\over R} + {\dot R^2\over R^2} - {3\over 4}{S^2\over R^4} = -\bar p + 2\eta{\dot S\over S}.\ee
where the dot indicates time differentiation. The expansion and shear scalars have usual definitions as given by Raychaudhuri \cite{11}
\be\label{2.10} \theta = {v^i}_{;i} \;\;\;\mathrm{and}\;\;\; \sigma^2 = {1\over 2}\sigma_{ij} \sigma^{ij},\ee
where
\be\label{2.11} \sigma_{ij} = v_{i;j} {1\over 2} \left(v_{i;a}v^a v_j + v_{j;a} v^a v_j\right) - {1\over 3}\theta(g_{ij} + v_i v_j).\ee
In view of the metric \eqref{2.1}, the expansion and shear scalars given by equation \eqref{2.10} can be written in a comoving coordinate system as
\be\label{2.12} \theta = 2{\dot R\over R} + {\dot S\over S},\ee
and
\be\label{2.13} \sigma^2 = {1\over 3}\left({\dot R\over R} - {\dot S\over S}\right)^2.\ee
As a consequence of the Bianchi identity we have
\be \label{2.14} \dot \rho = - (\rho + p)\theta + \zeta \theta^2 + 4\eta \sigma^2.\ee
The Raychaudhuri equation [12] can now be written
\be\label{2.15} \dot \theta = -{1\over 3}\theta^2 - 2\sigma^2 + R_{ij} v^i v^j,\ee
where
\be \label{2.16} R_{ij} v^i v^j = -{1\over 2}\big[\rho + 3(p-\zeta\theta)\big].\ee
It is clear from the above that the Hawking-Penrose energy condition \cite{13} is satisfied when $R_{ij} v^i v^j \le 0$, which yields in our case $\rho + 3(p - \zeta\theta)\ge 0$. So in a contracting model ($\theta < 0$), the energy condition is always satisfied so long as we demand positive magnitudes for fluid density and pressure from physical considerations. In an expanding model, however, the energy condition is satisfied provided that $\rho + 3p \ge 3\zeta \theta$. One should further note that if the energy condition is satisfied, that is, $R_{ij} v^i v^j \le 0$, we have in view of equation \eqref{2.15}, $\dot \theta < 0$, which shows that no bounce can take place from a minimum volume.\\

Now we have a system of three independent equations \eqref{2.7} - \eqref{2.9} with six unknown quantities, viz., $R,~ S,~ \rho,~ p,~ \eta, ~\mathrm{and} ~\zeta$. Hence we assume the following three appropriate relations among these variables in order to obtain solutions of these equations. One of them is the barotropic equation of state connecting the density and pressure by,

\be \label{2.17a} p = \epsilon \rho,~~~~\mathrm{where},~~~~0 < \epsilon \le 1.\ee
The other two relations are the linear relationships between $\rho,~ \theta^2~ \mathrm{and}~ \sigma^2$, that is,
\be \label{2.17b} {\rho\over \theta^2} = {1\over 3} {C_1}^2,\ee
and
\be \label{2.17c} {\sigma^2\over \theta^2} = {1\over 3} {C_2}^2,\ee
where $C_1$ and $C_2$ are two constants. Now substituting equations \eqref{2.12} and \eqref{2.13} in equation \eqref{2.17c}, one obtains the following relation:
\be \label{2.18} \left({C_2}^2 - 1\right){\dot S^2\over S^2}+ 2\left(2C_2^2 + 1\right){\dot R\over R} {\dot S\over S} + \left(4{C_2}^2 - 1\right){\dot R^2\over R^2} = 0.\ee
We have to exclude the value ${C_2}^2 = 1$, because in that case we get immediately from equations \eqref{2.7} and \eqref{2.18} the result $\rho < 0$ emerges, which is un-physical. So for ${C_1}^2 \ne 1$, the solution of the quadratic equation \eqref{2.18} yields

\be\label{2.19a} {\dot S\over S} = \lambda{\dot R\over R},\ee
which on integration gives
\be\label{2.19} S = R^\lambda.\ee
In the above $\lambda = ({C_2}^2- 1)^{-1}[-(2{C_2}^2 + 1) \pm 3C_2]$ and the arbitrary constant appearing in the process of integration has been absorbed into $S$ (or $R$) without any loss of generality. Using equation \eqref{2.19} the field equation \eqref{2.7} to \eqref{2.9} and the equations for expansion and shear scalars, viz., equations \eqref{2.12} and \eqref{2.13}, respectively, can be reduced to

\be\label{2.20a}(2\lambda + 1){\dot R^2\over R^2} - {1\over 4}R^{2(\lambda - 2)} = \rho,\ee
\be\label{2.20b} (\lambda+1){\ddot R\over R} + \lambda^2{\dot R^2\over R^2}+ {1\over 4}R^{2(\lambda - 2)} = \bar p + 2\eta {\dot R\over R},\ee
\be\label{2.20c} 2{\ddot R\over R} + {\dot R^2\over R^2} - {3\over 4}R^{2(\lambda - 2)} = -\bar p + 2\eta \lambda{\dot R\over R},\ee
\be\label{2.20d} \theta = (2+\lambda){\dot R\over R},\ee
\be\label{2.20e} \sigma^2 = {1\over 3}(1-\lambda)^2{\dot R^2\over R^2}.\ee
When $\sigma^2 =0$, the shear vanishes and we have from equation \eqref{2.13} ${\dot R\over R} = {\dot S\over S}$, so that the motion is isotropic. Since the present paper considers an anisotropic model with non-vanishing shear at any arbitrary time, we have
$\sigma^2 \ne 0$. This leads in our case to the conclusion that
\be \label{2.21} \lambda \ne 1.\ee
Again using equations \eqref{2.20d} and \eqref{2.20e} in equation \eqref{2.17c}, we find
\be \label{2.22} {C_2}^2 = (2 + \lambda)^{-2}(1 - \lambda)^2.\ee
Since ${C_2}^2 \ne 1$, we have from equation \eqref{2.22}
\be \label{2.23}\lambda \ne -{1\over 2}.\ee
Further, in view of equations \eqref{2.17b} and \eqref{2.20d}, one finds
\be \label{2.24} \rho = {1\over 3} {C_1}^2(2 + \lambda)^2 {\dot R^2 \over R^2},\ee
which together with equation \eqref{2.20a} yields
\be \label{2.25} R^{2(\lambda - 2)} = 4\left[(2\lambda +1) - {1\over 3}{C_1}^2 (2 + \lambda)^2\right]{\dot R^2\over R^2},\ee
from which it follows that, one must have at least
\be \label{2.26a} \lambda > - {1\over 2},\ee
and
\be\label{2.26b} {1\over 3} {C_1}^2 < {2\lambda + 1\over (2 + \lambda)^2}.\ee
We next make use of equation \eqref{2.25} to write equations \eqref{2.20b} and \eqref{2.20c} in the following forms, respectively:
\be\label{2.27a} -\bar p + 2\eta {\dot R\over R} = \left[2\lambda(1+\lambda) - {1\over 3}{C_1}^2(2 + \lambda)^2\right]{\dot R^2\over R^2},\ee
and
\be\label{2.27b} -\bar p + 2\eta\lambda {\dot R\over R} = \left[{C_1}^2(2+\lambda)^2 - 4(1 + \lambda)\right]{\dot R^2\over R^2},\ee
Subtracting equation \eqref{2.27b} from equation \eqref{2.27a} one readily obtains
\be\label{2.27c}  2\eta(1 - \lambda) {\dot R\over R} = 2(2+\lambda)\left[(1+\lambda) - {2\over 3}{C_1}^2(2 + \lambda)\right]{\dot R^2\over R^2},\ee
which in turn yields
\be\label{2.28} \eta = \left({2 + \lambda\over 1 - \lambda}\right)
\left[(1 + \lambda) - {2\over 3}{C_1}^2(2 + \lambda)\right]{\dot R\over R}.\ee
Using equations \eqref{2.10}, \eqref{2.20d}, and \eqref{2.28} one obtains from equation \eqref{2.3} the following form of the effective pressure:
\be \label{2.29a} \bar p = (p - \zeta \theta) + {2\over 3}\left({2+\lambda\over 1 - \lambda}\right)\left[(1 + \lambda)(2 + \lambda) - {2\over 3}{C_1}^2(2 + \lambda)^2\right]{\dot R^2\over R^2}.\ee
Again using equation \eqref{2.28} in equation \eqref{2.27a} one further obtains
\be \label{2.29b}\bar p = \left({1\over 1 - \lambda}\right)\left[2(1 + \lambda)(2 + \lambda^2) - {1\over 3}{C_1}^2(3 + \lambda)(2 + \lambda)^2\right]{\dot R^2\over R^2}.\ee
It then follows from the above two relations \eqref{2.29a} and \eqref{2.29b} that
\be \label{2.30} (p - \zeta \theta) = {1\over 3}\left[4(1-\lambda^2) - {1\over 3}{C_1}^2(2+\lambda)^2\right]{\dot R^2\over R^2}.\ee
Now in view equations \eqref{2.24} and \eqref{2.30} the Hawking-Penrose energy condition
$\rho + 3(p - \zeta \theta) \ge 0$, obtained previously, is satisfied if the following relation holds good. It is
\be \label{2.31a} 4(1 - \lambda^2){\dot R^2\over R^2}\ge 0,\ee
which in turn demands $\lambda^2 < 1$, because equation \eqref{2.21} excludes the possibility of $\lambda^2$ being unity. It then follows that $\lambda$ must be less than unity. Further using equations \eqref{2.20d}, \eqref{2.20e}, \eqref{2.24}, \eqref{2.30}, and \eqref{2.16} in the Raychaudhuri equation \eqref{2.15}, one gets
\be\label{2.31} {\ddot R\over R} = (\lambda - 1){\dot R^2\over R^2},\ee
which, in view of the fact that $\lambda < 1$ leads to the following result,
\be \label{2.31b} \ddot R < 0,\ee
which implies that the dimension $R$ cannot have any minimum. If it has at all any turning point, it can only be a maximum. Thus the singularity of zero volume cannot be avoided.\\

All the results obtained so far are independent of any specific equation of state. Now using the barotropic equation of state given by equation \eqref{2.17a} in equation \eqref{2.24} we obtain
\be\label{2.32} \rho = {1\over 3}\epsilon {C_1}^2 (2 + \lambda)^2{\dot R^2\over R^2}.\ee
Using equations \eqref{2.20d} and \eqref{2.32} in equation \eqref{2.30} we obtain for the bulk viscosity coefficient
\be\label{2.33} \zeta = {1\over 3(2+\lambda)}\left[{1\over 3}{C_1}^2 (1 + 3\epsilon)(2+\lambda)^2 -4(1-\lambda)^2\right]{\dot R\over R}.\ee
The physical requirement that both $\eta$ and $\zeta$ are positive therefore yields in view of equations \eqref{2.28} and \eqref{2.33} the following two inequalities for the expanding and contracting models. For expansion
\be\label{2.34a}{4(1-\lambda^2)\over (1+3\epsilon)(2+\lambda)^2} < {1\over 3}{C_1}^2 < {1+\lambda\over 2(2+\lambda)},\ee
and for contraction
\be\label{2.34b}{1+\lambda\over 2(2+\lambda)} < {1\over 3}{C_1}1^2 < {4(1-\lambda^2)\over (1+3\epsilon)(2+\lambda)^2}.\ee
One should remember that the condition (26b), that is, ${1\over 3}{C_1}^2 < {2\lambda + 1\over(2+\lambda)^2}$, however, in general holds irrespective of any particular model. It is to be noted that both the equations \eqref{2.34a} and \eqref{2.34b} cannot be simultaneously satisfied for the same set of $\epsilon,~\lambda,~ \mathrm{and}~ {C_1}$. Hence unlike the perfect fluid case one finds either an expanding model or a contracting model according to whether $\epsilon,~\lambda,~ \mathrm{and}~ {C_1}$ satisfy conditions \eqref{2.34a} or \eqref{2.34b}, respectively. For an expanding model we have from condition (34a) the following restrictions:
\be\label{2.34c} {2\over 3}{C_1}^2 (2+\lambda)^2 < (1+\lambda)(2+\lambda),\ee
and also
\be\label{2.34d} {2\over 3}{C_1}^2 (2+\lambda)^2 (1 + 3\epsilon) > 8(1-\lambda^2).\ee
The above two conditions when combined imply the restriction
\be\label{2.34e} \lambda > 2\left({1-\epsilon\over \epsilon + 3}\right),\ee
which in turn allows only positive values to the parameter $\lambda$. This restriction
on $\lambda$ is again consistent with the condition ${1+\lambda\over 2(2+\lambda)} < {2\lambda+1\over (2+\lambda)^2}$, so that in this case the condition \eqref{2.26b} is already satisfied. Thus the limits of $\lambda$ can be expressed in the form
\be\label{2.34f} 0 \le 2{\left(1-\epsilon\over \epsilon + 3\right)} < \lambda < 1,~~~\mathrm{for}~~~0 < \epsilon \le 1.\ee
For contraction, on the other hand, we use the condition \eqref{2.26b} in the condition \eqref{2.34b} to get
\be\label{2.34g} {1+\lambda\over 3(2+\lambda)} <{1\over 3}{C_1}^2 < {2\lambda + 1\over (2+\lambda)^2},\ee
which yields the relation
\be\label{2.34h} \lambda(\lambda -1) < 0.\ee
This relation, in view of the fact that $\lambda < 1$, which was proved earlier,
establishes that in this case $\lambda$ must also be positive. Thus in this case the
restriction on $\lambda$ can be expressed as:
\be\label{2.34i} 0 < \lambda < 2{\left(1-\epsilon\over \epsilon + 3\right)} < 1,~~~\mathrm{for}~~~0<\epsilon < 1.\ee
One should note here that the extreme case of a stiff fluid is not allowed
here. Now the equation \eqref{2.25} can be written as:
\be\label{2.34j} R^{(1-\lambda)}\dot R = \pm {1\over 2}\left[(2\lambda + 1)-{1\over 3}{C_1}^2(2+\lambda)^2\right]^{-{1\over 2}},\ee
where the positive and negative signs are corresponding to expansion $(\dot R > 0)$ and contraction $(\dot R < 0$), respectively. If now we consider the expanding model, the above equation can be integrated to yield
\be\label{2.35a} R^{(2-\lambda)} = {1\over 2}(2-\lambda)\left[(2\lambda + 1)-{1\over 3}{C_1}^2(2+\lambda)^2\right]^{-{1\over 2}}(t + t_1),\ee
where the constant of integration $t_1$ has been fixed by choosing $t= -t_1$ as the epoch when $R = 0$ and hence the proper volume, which is proportional to $R^2 S$ is also zero. The contracting model on the other hand is given by:
\be\label{2.35b} R^{(2-\lambda)} = {1\over 2}(2-\lambda)\left[(2\lambda + 1)-{1\over 3}{C_1}^2(2+\lambda)^2\right]^{-{1\over 2}}(t_2 - t).\ee
Here again the constant of integration $t_2$ is fixed by choosing $t = t_2$ as the
epoch when $R$ and hence the proper volume vanish. Now writing
\be\label{2.36} R_0 = \left\{{1\over 2}(2-\lambda)\left[(2\lambda + 1)-{1\over 3}C_1^2(2+\lambda)^2\right]^{-{1\over 2}}\right\}^{1\over 2-\lambda}.\ee
One can now express $R$ as:
\be\label{2.37a} R = R_0 (t_1+t)^{\left(1\over 2-\lambda\right)},\ee
and
\be\label{2.37b} R = R_0 (t_2-t)^{\left(1\over 2-\lambda\right)},\ee
for expanding and contracting models, respectively. Accordingly in view of equations \eqref{2.37a} and \eqref{2.37b} $S$ can be calculated from equation \eqref{2.19} as
\be\label{2.38a} S = {R_0}^\lambda (t_1+t)^{\left(\lambda\over 2-\lambda\right)},\ee
and
\be\label{2.38b} S = {R_0}^\lambda (t_2-t)^{\left(\lambda\over 2-\lambda\right)},\ee
for expanding and contracting models, respectively. The metric \eqref{2.1} can now be expressed correspondingly in the forms:
\be\label{2.39a} ds^2 = -dt^2 + {R_0}^{2\lambda}(t_1+t)^{\left(2\lambda\over 2-\lambda\right)}(dx+zdy)^2 +{R_0}^2(t_1+t)^{\left(2\over 2-\lambda\right)}(dy^2+dz^2),~~~-t_1\le t \le\infty,\ee
and
\be\label{2.39b} ds^2 = -dt^2 + {R_0}^{2\lambda}(t_2-t)^{\left(2\lambda\over 2-\lambda\right)}(dx+zdy)^2 +{R_0}^2(t_2-t)^{\left(2\over 2-\lambda\right)}(dy^2+dz^2),~~~~0\le t \le t_2.\ee
Finally using equations \eqref{2.37a} and \eqref{2.37b} in equations \eqref{2.20d}, \eqref{2.20e}, \eqref{2.24}, \eqref{2.32}, \eqref{2.28}, and \eqref{2.33}, we obtain for an expanding model:

\be\label{2.40a} \theta = \left({2+\lambda\over 2-\lambda}\right)\left(1\over t_1+t\right),\ee
\be\label{2.40b}  \sigma^2 ={1\over 3}\left({1-\lambda\over 2-\lambda}\right)^2\left(1\over t_1+t\right)^2,\ee
\be\label{2.40c} \rho = {1\over 3}{C_1}^2 \left({2+\lambda\over 2-\lambda}\right)^2\left(1\over t_1+t\right)^2,\ee
\be\label{2.40d}  p = {1\over 3}\epsilon {C_1}^2  \left({2+\lambda\over 2-\lambda}\right)^2\left(1\over t_1+t\right)^2,\ee
\be\label{2.40e} \eta = {2+\lambda\over (2-\lambda)(1-\lambda)}\left[(1+\lambda) - {2\over 3}{C_1}^2 (2 + \lambda)\right]\left(1\over t_1+t\right),\ee
\be\label{2.40f} \zeta = {1\over 3(2-\lambda)}\left[{1\over 3}{C_1}^2(1+3\epsilon)(2+\lambda) - {4(1-\lambda^2)\over 2+\lambda}\right]\left(1\over t_1+t\right).\ee
For a contracting model
\be\label{2.40g}\theta = - \left[{2+\lambda\over 2 -\lambda}\right]\left[{1\over t_2 - t}\right].\ee The rest of the scalars can be obtained by replacing $(t_l + t)$ in equations \eqref{2.40b} to \eqref{2.40f} by $(t_2-t)$. Thus equations \eqref{2.37a}, \eqref{2.37b}, \eqref{2.38a}, \eqref{2.38b} and \eqref{2.40c} - \eqref{2.40f} and those for the corresponding contracting case constitute the complete set of solutions of the field equations \eqref{2.7} - \eqref{2.9}. The metric \eqref{2.39a} is singular at $t = -t_l$ in the case of expanding model, while the metric \eqref{2.39b} is singular at $t = t_2$ in the case of the contracting model. Since $\lambda$ is positive, both $R$ and $S$ approach vanishingly small magnitudes near the singularity. Now the proper volume is proportional to $R^2 S$, which again is equal to $R^{2+\lambda}$. So proper volume tends to zero as $R \rightarrow 0$. We obtain a point type singularity at this limit. But since $\lambda < 1$ and $S = R^\lambda$, we observe that $S$ approaches zero at different rate than $R$. Again it follows from equations \eqref{2.40a} - \eqref{2.40f} that all the kinematical quantities and the fluid variables such as ($\theta, \sigma^2, \rho, p, \eta, \zeta$ etc.), blow up in the limit $R \rightarrow 0$ developing a singularity. In an expanding model the viscosity coefficients which have infinitely large magnitudes near the initial singular state monotonically decrease with time and assume vanishingly small values at the final stage of evolution. In the case of a contracting model, however, these coefficients start from finite magnitudes at $t = 0$ and monotonically increase with time, assuming infinitely large values at the final stage of collapse. The magnitude of the shear as well as the expansion rate diminishes in the course of expansion. Further, it is interesting to note that the solutions given by equations \eqref{2.37a}, \eqref{2.37b} and \eqref{2.38a}, \eqref{2.38b} for LRS spatially homogeneous Bianchi type-II model are self-similar in the sense that they admit homothetic motion. This can be shown explicitly in the following way. If $b$ is a constant, the transformations $t\rightarrow b^{2-\lambda}t,~x \rightarrow b^{2(1-\lambda)}x,~y \rightarrow b^{1-\lambda}y,~\mathrm{and}~z\rightarrow b^{1-\lambda}z$, lead to the metric $ds^2 = b^{2(2-\lambda)}ds^2$, which characterizes a homothetic motion. Here the time coordinate $t$ is written for $(t + t_1)$.\\

Now the hydrodynamic equation \eqref{2.14} is related to the law of increase
of entropy for a given energy dissipation. If we define the entropy density $\Sigma$ (see Belinskii and Khalatnikov \cite{4}) as
\be\label{2.41a} \Sigma = \Sigma_0 \exp{\left[\int {d\rho\over \rho + p(\rho)}\right]},\ee
where, $\Sigma_0$ is a constant. The equation \eqref{2.14} then can be written, in view of equation \eqref{2.12}, as
\be\label{2.41b} {d\over dt}\left[\ln{(\Sigma R^2 S)}\right] = {\zeta \theta^2 + 4\eta\sigma^2\over p+\rho}.\ee
$\Sigma R^2 S$ may be said to be the rate of change of entropy with time and is
clearly seen to be greater than zero so long as the physical quantities such
as $\rho,~p,~\zeta,~\mathrm{and}~\eta$ remain positive. This signifies the law of increase of total entropy in the future course of evolution of the universe. Now substituting the expressions for $\theta,~\sigma^2,~\rho,~p,~\eta,~\mathrm{and}~\zeta $ from equations \eqref{2.20d}, \eqref{2.20e}, \eqref{2.24}, \eqref{2.32} and \eqref{2.33} in the above equation, we find
\be\label{2.42a} {d\over dt}\left[\ln{(\Sigma R^2 S)}\right] = k{\dot R\over R},\ee
with
\be\label{2.42b} k =(2-\lambda)\left({2\lambda\over 2-\lambda} + {\epsilon-1\over \epsilon+1}\right).\ee
It may be readily seen that $k$ and $\dot R\over R$ must have the same sign by virtue of the law of increase of entropy and this puts some restrictions on the relative magnitudes of $\lambda$ and $\epsilon$ for two separate cases of expansion and contraction of the models. The conclusions, however, do not contradict the limits already obtained. Now equation \eqref{2.42a} may be integrated to yield,
\be\label{2.43} \Sigma R^2 S = A R^k,\ee
where $A$ is a positive definite arbitrary constant. Equation \eqref{2.43} implies that in an expanding model $(k > 0)$ the entropy increases with the increase of $R$, whereas in a collapsing model $(k < 0)$ the entropy increases with the decrease of $R$. Further in view of equations \eqref{2.37a}, \eqref{2.37b} and \eqref{2.43} the entropy in an expanding model is given by
\be\label{2.44a} \Sigma R^2 S = A {R_0}^k(t_1 + t)^{k\over 2-\lambda},~~~k > 0,\ee
and that in a contracting model is given by
\be\label{2.44b} \Sigma R^2 S = A {R_0}^k(t_2 - t)^{k\over 2-\lambda},~~~k < 0.\ee
The time derivation of the total entropy is calculated from equation \eqref{2.44a} to yield
\be\label{2.444} {d\over dt}(\Sigma R^2 S) = {Ak\over 2-\lambda} {R_0}^k (t_1 + t)^{{k\over 2-\lambda}-1}.\ee
Clearly the time rate of entropy increase is indefinitely large at the initial
stage $t = -t_1$ provided ${k\over 2-\lambda}- 1 < 0$. Substituting the value of $k$ from equation \eqref{2.42b}, this condition is found to be equivalent to $\lambda < {2\over \epsilon + 2}$. So for $\lambda < {2\over 3}$, the above condition is fulfilled for the whole range $0 < \epsilon \le 1$. Thus within this limit for $\lambda$ the entropy generation process is very large at the initial instant compared to the subsequent stages.

\section{Special case.}
Having found the general solutions, we now move on to study some special case, particularly when bulk viscosity vanishes (case-1), and when both the bulk and the shear viscosity vanishes, ie. the case of perfect fluid (case-2). This will allow us to compare the effect of viscosity on the evolution of the universe.

\subsection{Case 1: $\zeta = 0$}
In the absence of the bulk viscosity the solutions \eqref{2.37a}, \eqref{2.37b} and \eqref{2.38a}, \eqref{2.38b} given earlier remain valid, but here the equation of state $p = \epsilon \rho$ is no longer an additional assumption, since it follows directly from the field equations. The equation \eqref{2.30} now reduces to
\be\label{2.45} p = {1\over 3} \left[4(1-\lambda^2) -{1\over 3} {C_1}^2(2+\lambda)^2\right]{\dot R^2\over R^2}.\ee
Using equation \eqref{2.24} in the above equation \eqref{2.45} we have
\be\label{2.46} p = \left[{4(1-\lambda^2)\over {C_1}^2(2+\lambda)^2} - {1\over 3}\right]\rho,\ee
which implies a barotropic equation of state $p = \epsilon\rho$, where
\be\label{2.47}\epsilon = \left[{4(1-\lambda^2)\over {C_1}^2(2+\lambda)^2} - {1\over 3}\right].\ee
For a physically meaningful fluid with a stiff fluid approximation in the
limit, $\epsilon$ has limits given by $0 < \epsilon \le 1$, so that we have
\be\label{2.48} 0 < \left[{4(1-\lambda^2)\over {C_1}^2(2+\lambda)^2} - {1\over 3}\right] \le 1.\ee
The inequality relation \eqref{2.48} immediately leads us to the result $1 - \lambda^2 > 0$, so that $\lambda^2 < 1$. Again, in view of equation \eqref{2.26a}, we remember that $\lambda$ is greater than at least $- {1\over 2}$, so that $1 + \lambda > 0$ and as a result $1 - \lambda > 0$. This
definitely establishes the fact that $\lambda$ can assume only values less than unity. From physical considerations we have to allow only positive values of $\eta$, so that in view of equation \eqref{2.28} we get for expansion the condition ${1\over 3} {C_1}^2 < {1 + \lambda \over 2(2 + \lambda)}$. This condition when considered together with the condition \eqref{2.48} demands
\be\label{2.49} {1-\lambda^2\over(2+\lambda)^2} < {1 +\lambda \over 2(2+\lambda)},\ee
which, since $1 + \lambda > 0$, finally requires $\lambda > 0$. So the allowed limits on $\lambda$ for an expanding model in this case are given by $0 < \lambda < 1.$ Again the positive value of $\eta$ in a contracting model demands ${1\over 3} {C_1}^2 > {1 + \lambda \over 2(2 + \lambda)}$, which when combined with \eqref{2.48} gives
\be\label{2.50} {4(1-\lambda^2)\over(2+\lambda)^2} > {1 +\lambda \over 2(2+\lambda)}.\ee
The condition \eqref{2.50} immediately requires ${2\over 3}$ to be the upper bound for $\lambda$. In general, we have from the condition \eqref{2.26b}, ${1\over 3} {C_1}^2 < {2\lambda +1 \over (2 + \lambda)^2}$, so that for contracting models the following relation is satisfied:
\be\label{2.51} {1 +\lambda \over 2(2+\lambda)} < {1\over 3} {C_1}^2 < {2\lambda +1 \over (2 + \lambda)^2},\ee
which in turn requires that $\lambda(\lambda -1) < 0$, that is, $0 < \lambda < 1$. The more stringent restrictions on $\lambda$ in a contracting model are, therefore, given by
\be\label{2.52} 0 < \lambda < {2\over 3}.\ee

\subsection{Case 2: $\eta = \zeta = 0$}
It is the case of a perfect fluid. Since $\eta = 0$, the equation \eqref{2.28} yields the condition
\be\label{2.53} {1\over 3}{C_1}^2 = {1+\lambda\over 2(2+\lambda)},\ee
and equation \eqref{2.30}, in view of $\zeta = 0$ in this case leads directly to
\be\label{2.54} p = {1\over 3}\left[4(1-\lambda^2) - {1\over 3}{C_1}^2 (2+\lambda)^2\right]{\dot R^2\over R^2}.\ee
Using equation \eqref{2.53} in the expressions \eqref{2.54} and \eqref{2.24} one obtains the following two relations:
\be\label{2.55} p = {1\over 2}(1+\lambda)(2-3\lambda){\dot R^2\over R^2}\ee
and
\be\label{2.56} \rho = {1\over 2}(1+\lambda)(2+\lambda){\dot R^2\over R^2}\ee
which relate the pressure and density in the following manner:
\be\label{2.57} p = \left[{2-3\lambda\over 2+\lambda}\right]\rho.\ee
This is evidently a barotropic form of the equation of state $p = \epsilon \rho$. Again equations \eqref{2.26b} and \eqref{2.53} together can be combined to yield
\be\label{2.57a} {2\lambda +1 \over (2+\lambda)^2} > {1+\lambda\over 2(2+\lambda)},\ee
which in turn demands $\lambda(\lambda-1) < 0$. The only possibility is that $\lambda$ must lie between zero and unity. Further since $\lambda$ is greater than zero, it excludes the limiting case for stiff fluid with $p = \rho$ as is evident from the equation \eqref{2.57}. The physical requirement that $0 \le p < \rho$ can now be satisfied provided under the condition:
\be\label{2.58} 0 < \lambda \le {2\over 3}.\ee
Further, the Hawking-Penrose energy condition, which demands that $\lambda^2 \le 1$ is satisfied if equation \eqref{2.58} holds. Now from equations \eqref{2.17b} and \eqref{2.53} we find that
\be\label{2.58a} {\rho\over \theta^2} = {1\over 3}{C_1}^2 = {1+\lambda\over 2(2+\lambda)}.\ee
It is not difficult to show from the above relation that ${\rho\over \theta^2}$ is a monotonically increasing function of $\lambda$ and as such the minimum and
maximum of this quantity will depend on the minimum and maximum allowable values for $\lambda$. Thus we get,
\be\label{2.59} {1\over 4} < {\rho\over \theta^2} \le {5\over 16}.\ee
Further, combining the equations \eqref{2.17c} and \eqref{2.22} we moreover obtain
\be\label{2.59a} {\sigma^2\over \theta^2} = {1\over 3}{C_2}^2 = {(1-\lambda)^2\over 3(2+\lambda)^2}.\ee
The quantity ${\sigma^2\over \theta^2}$ is a monotonically decreasing function of $\lambda$, and thus by the similar consideration we get its limits,
\be \label{2.60} {1\over 192} \le {\sigma^2\over \theta^2} < {1\over 12}.\ee
Now writing $\gamma = {2(2-\lambda)\over (2+\lambda)}$, we have, in view of equation \eqref{2.58} the limits on $\gamma$ given by
\be \label{2.61} 1\le \gamma < 2.\ee
One can express the equation of state \eqref{2.57} in the form
\be \label{2.62} p =  (\gamma - 1)\rho.\ee
The solutions of $R$ and $S$ given in the equations \eqref{2.37a}, \eqref{2.37b} and \eqref{2.38a}, \eqref{2.38b} for expanding and contracting models, can be explicitly written, expressing the powers in terms of $\gamma$ in place of $\lambda$ as,
\be \label{2.63a} R^2 = \left[{2\gamma(t_1 + t)\over [(3\gamma-2)(2-\gamma)]^{1\over 2}}\right]^{2+\gamma\over 2\gamma},\ee
\be \label{2.63b} R^2 = \left[ {2\gamma(t_2 - t)\over [(3\gamma-2)(2-\gamma)]^{1\over 2}}\right]^{2+\gamma\over 2\gamma},\ee
\be \label{2.64a} S^2 = \left[{2\gamma(t_1 + t)\over [(3\gamma-2)(2-\gamma)]^{1\over 2}} \right]^{2-\gamma\over \gamma},\ee
\be \label{2.64b} S^2 = \left[{2\gamma(t_2 - t)\over [(3\gamma-2)(2-\gamma)]^{1\over 2}} \right]^{2-\gamma\over \gamma},\ee
respectively. One can recognize the solutions \eqref{2.63a} and \eqref{2.64a} for an expanding universe as that mentioned by Collins and Stewart \cite{14}. Since ${2+\gamma\over 2\gamma}$ and ${2-\gamma\over \gamma}$ are both positive, so near the initial singularity, when $R\rightarrow 0$, $S$ also tends to zero. But since $R^2 \sim (t_1 + t)^{2+\gamma\over 2\gamma}$ and $S^2 \sim (t_1 + t)^{2-\gamma \over \gamma}$, so $R^2$ goes to zero at a faster rate than $S^2$ does. Hence the nature of the singularity is the same as that observed in the cases of viscous fluid.\\

It is evident from the above analysis that the same set of solutions
\eqref{2.36}, \eqref{2.37a}, \eqref{2.37b}, \eqref{2.38a} and \eqref{2.38b} formally represent both viscous and perfect fluid except for ${C_1}^2$, which is being different in the two cases. When ${C_1}^2$ assumes a limiting value ${C_1}^2 = {1+\lambda\over 2(2+\lambda)} $ as given in equation \eqref{2.53} we observe the fluid to behave like that of a perfect fluid with the equation of state $p = \epsilon\rho$, where $\epsilon = {2-3\lambda\over 2+\lambda}$. Other values of ${C_1}^2$ and $\epsilon$ may represent viscous fluids under the restrictions mentioned in the text.\\

\noindent
\textbf{Acknowledgements:}
The authors would like to thank the referee for some valuable suggestions and also the U.G.C. (India) for financial support.

\end{document}